\documentstyle[13lomcon,cite,epsfig]{article}

\begin{document}

\title{UPPER BOUND ON THE LIGHTEST NEUTRALINO MASS \\
IN THE MINIMAL NON--MINIMAL SUPERSYMMETRIC STANDARD MODEL
}

\author{S.~Hesselbach\,,\quad G.~Moortgat-Pick}
\address{IPPP, University of Durham, Durham, DH1 3LE, U.K.\\~~~}

\author{D.~J.~Miller\,,\quad R.~Nevzorov 
\footnote{On leave of absence from the Theory Department, ITEP, Moscow, Russia.}
\footnote{e-mail: r.nevzorov@physics.gla.ac.uk}}
\address{Department of Physics and Astronomy, University of Glasgow, \\
Glasgow, G12 8QQ, U.K.\\~~~}

\author{M.~Trusov}
\address{Theory Department, ITEP, Moscow, 117218, Russia\\~~~}


\maketitle\abstracts{We consider the neutralino sector in the 
Minimal Non--minimal Supersymmetric Standard Model (MNSSM). We argue
that there exists a theoretical upper bound on the lightest neutralino 
mass in the MNSSM. An approximate solution for the mass of the 
lightest neutralino is obtained.}

Supersymmetric (SUSY) models provide an elegant explanation for the
dark matter energy density observed in the Universe. To prevent rapid 
proton decay in SUSY models the invariance of the Lagrangian under R--parity 
transformations is usually imposed. As a consequence the lightest 
supersymmetric particle (LSP) is absolutely stable and can play the role of 
dark matter. In most supersymmetric scenarios the LSP is the lightest
neutralino, which provides the correct relic abundance of dark matter
if it has a mass of $\mathcal{O}$(100~GeV).

In this article we explore the neutralino sector in the framework of the 
simplest extension of the minimal SUSY model (MSSM) --- the Minimal Non--minimal 
Supersymmetric Standard Model (MNSSM). The superpotential of the MNSSM
can be written as follows
\cite{Panagiotakopoulos:1999ah,Panagiotakopoulos:2000wp,Dedes:2000jp}
\begin{equation}
W_{MNSSM}=\lambda \hat{S}(\hat{H}_d \epsilon \hat{H}_u)+\xi \hat{S}+W_{MSSM}(\mu=0)\,,
\label{2}
\end{equation}
where $W_{MSSM}(\mu=0)$ is the superpotential of the MSSM without 
$\mu$-term. The superpotential (\ref{2}) does not contain any bilinear terms
avoiding the $\mu$--problem. At the same time quadratically divergent 
tadpole contributions can be suppressed in the considered model so that $\xi\le 
(\mbox{TeV})^2$ \cite{Panagiotakopoulos:1999ah,Panagiotakopoulos:2000wp}.
At the electroweak (EW) scale the superfield $\hat{S}$ gets a non--zero vacuum 
expectation value ($\langle S \rangle =s/\sqrt{2}$) and an effective $\mu$-term 
($\mu_{eff}=\lambda s/\sqrt{2}$) is automatically generated.

The neutralino sector of the MNSSM is formed by the superpartners of
the neutral gauge and Higgs bosons.
In the field basis $(\tilde{B},\,\tilde{W}_3,\,\tilde{H}^0_d,\,\tilde{H}^0_u,\,\tilde{S})$
the neutralino mass matrix reads
\begin{equation}
\begin{array}{l}
M_{\tilde{\chi}^0}=\\
\left(
\begin{array}{ccccc}
M_1                  & 0                  & -M_Z s_W c_{\beta}   & M_Z s_W s_{\beta}  & 0 \\[2mm]
0                    & M_2                & M_Z c_W c_{\beta}    & -M_Z c_W s_{\beta} & 0 \\[2mm]
-M_Z s_W c_{\beta}   & M_Z c_W c_{\beta}  &  0                   & -\mu_{eff}         & -\displaystyle\frac{\lambda v}{\sqrt{2}} s_{\beta} \\[2mm]
M_Z s_W s_{\beta}    & -M_Z c_W s_{\beta} & -\mu_{eff}           &  0                 & -\displaystyle\frac{\lambda v}{\sqrt{2}} c_{\beta} \\[2mm]
0                    & 0                  & -\displaystyle\frac{\lambda v}{\sqrt{2}} s_{\beta}  & -\displaystyle\frac{\lambda v}{\sqrt{2}} c_{\beta} 
& 0
\end{array}
\right),
\end{array}
\label{1}
\end{equation}
\noindent
where $M_1$ and $M_2$ are the $U(1)_Y$ and $SU(2)$ gaugino masses while $s_W=\sin\theta_W$, $c_W=\cos\theta_W$, $s_{\beta}=\sin\beta$, 
$c_{\beta}=\cos\beta$ and $\mu_{eff}=\lambda s/\sqrt{2}$. Here we introduce $\tan\beta=v_2/v_1$ and 
$v=\sqrt{v_1^2+v_2^2}=246\,\mbox{GeV}$, where $v_1$ and $v_2$ are the vacuum expectation values of  
the Higgs doublets fields $H_d$ and $H_u$, respectively. From Eq.(\ref{1}) one can easily see that the neutralino spectrum 
in the MNSSM may be parametrised in terms of
\begin{equation}
\lambda\,,\qquad \mu_{eff}\,,\qquad \tan\beta\,, \qquad M_1\,,\qquad M_2\,.
\label{3}
\end{equation}
In supergravity models with uniform gaugino masses at the Grand Unification scale the renormalisation 
group flow yields a relationship between $M_1$ and $M_2$ at the EW scale, i.e. $ M_1\simeq 0.5 M_2$.
The chargino masses in the MNSSM are also defined by the mass parameters $M_2$ and $\mu_{eff}$. 
LEP searches for SUSY particles set a lower limit on the chargino masses of about $100\,\mbox{GeV}$
restricting the allowed interval of $|M_2|$ and $|\mu_{eff}|$ above $90-100\,\mbox{GeV}$.

In contrast with the MSSM the allowed range of the mass of the lightest neutralino in the MNSSM is limited.
In Fig.~1 we plot the lightest neutralino mass $|m_{\chi^0_1}|$ in the MSSM and MNSSM as a function of $M_2$ 
for different values of $\mu_{eff}$. From Fig.~1 it becomes clear that the absolute value of the mass of the 
lightest neutralino in the MSSM grows when $|M_2|$ and $|\mu_{eff}|$ increase while in the MNSSM 
the maximum value of $|m_{\chi^0_1}|$ reduces with increasing $|M_2|$ and $|\mu_{eff}|$. In order to find the upper bound on 
$|m_{\chi^0_1}|$ it is convenient to consider the matrix $M_{\tilde{\chi}^0} M^{\dagger}_{\tilde{\chi}^0}$ 
whose eigenvalues are equal to the absolute values of the neutralino masses squared. In the basis 
$\left(\tilde{B},\,\tilde{W}_3,\,-\tilde{H}^0_d s_{\beta}+\tilde{H}^0_u c_{\beta},\, \tilde{H}^0_d c_{\beta}+\tilde{H}^0_u s_{\beta},
\,\tilde{S}\right)$ the bottom-right $2\times 2$ block of $M_{\tilde{\chi}^0} M^{\dagger}_{\tilde{\chi}^0}$ takes the form
\cite{neutralino} 
\begin{equation}
\left(
\begin{array}{c@{\qquad\qquad}c}
|\mu_{eff}|^2+\tilde{\sigma}^2           & \nu^{*}\mu_{eff} \\
 \nu\mu^{*}_{eff}                             & |\nu|^2
\end{array}
\right),
\label{4}
\end{equation}
where $\tilde{\sigma}^2=M_Z^2\cos^2 2\beta+|\nu|^2\sin^2 2\beta,\,\nu=\lambda v/\sqrt{2}$.
Since the minimal eigenvalue of any hermitian matrix is less than its smallest diagonal element
the lightest neutralino in the MNSSM is limited from above by the bottom--right diagonal
entry of matrix (\ref{4}), i.e. $|m_{\chi^0_1}|\le |\nu|$. At the same time since we can
always choose the field basis in such a way that the $2\times 2$ submatrix (\ref{4}) 
becomes diagonal its minimal eigenvalue $\mu^2_0$ also restricts the allowed interval of $|m_{\chi^0_1}|$, i.e.
\begin{equation}
\begin{array}{c}
|m_{\chi^0_1}|^2\le \mu_{0}^2=\displaystyle\frac{1}{2}\biggl[|\mu_{eff}|^2+\tilde{\sigma}^2+|\nu|^2\quad\qquad\qquad\qquad\qquad\\[1mm]
\quad\qquad\qquad\qquad\qquad
-\sqrt{\biggl(|\mu_{eff}|^2+\tilde{\sigma}^2+|\nu|^2\biggr)^2-4|\nu|^2\tilde{\sigma}^2}\biggr]\,.
\end{array}
\label{5}
\end{equation}
The value of $\mu_0$ reduces with increasing $|\mu_{eff}|$. It reaches its maximum value, i.e.
$\mu_0^2=\min\biggl\{\tilde{\sigma}^2,\,|\nu|^2 \biggr\}$, when $\mu_{eff}\to 0$. Taking into account 
the restriction on the effective $\mu$--term coming from LEP searches and the theoretical upper bound on the Yukawa
coupling $\lambda$ which is caused by the requirement of the validity of perturbation theory up to
the Grand Unification scale ($\lambda<0.7$) we find that $|m_{\chi^0_1}|$ does not exceed $80-85\,\mbox{GeV}$
at tree level \cite{neutralino,Hesselbach:2007ta}.
\begin{figure}[hbtp]
\begin{center}
\hspace*{-5.5cm}{$|m_{\chi^0_1}|$}\hspace*{5.5cm}{$|m_{\chi^0_1}|$}\\
\mbox{
\epsfig{file=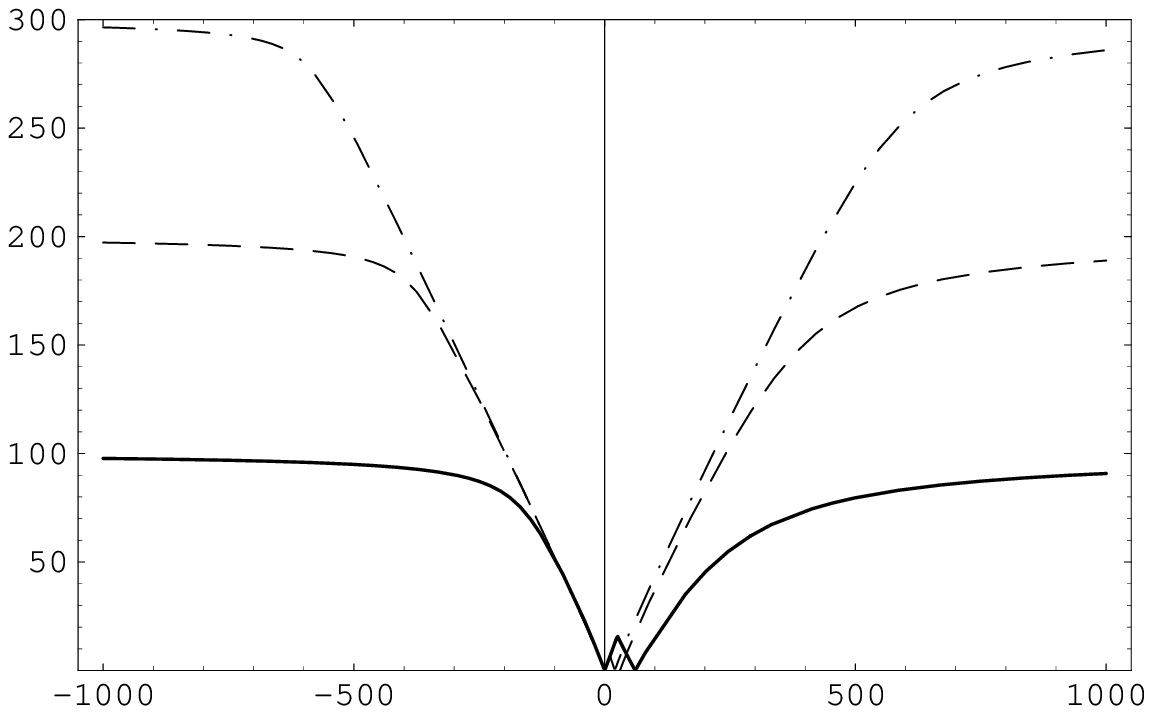,height=3.7 cm}
\epsfig{file=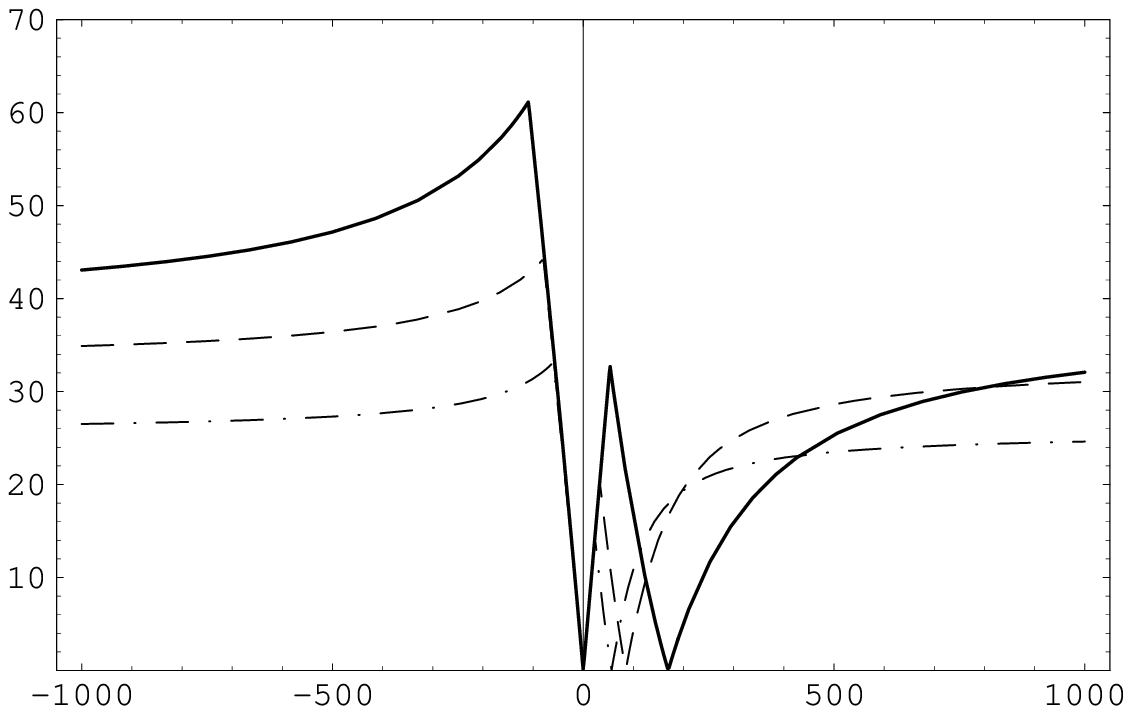,height=3.7 cm}
}\\
\hspace*{0.5cm}{$M_2$}\hspace*{5.5cm}{$M_2$}\\[1mm]
\hspace*{0.5cm}{\bf (a)}\hspace*{5.5cm}{\bf (b) }
\end{center}
\caption{Lightest neutralino mass versus $M_2$ in the {\it (a)} MSSM and {\it (b)} MNSSM for
$\tan\beta=3$, $\lambda=0.7$, $M_1=0.5 M_1$. Solid, dashed and dash--dotted lines correspond 
to $\mu_{eff}=100\,\mbox{GeV}, 200\,\mbox{GeV}$ and $300\,\mbox{GeV}$, respectively.} 
\end{figure}

Here it is worth to notice that at large values of $\mu_{eff}$ the allowed interval of the lightest 
neutralino mass shrinks drastically. Indeed, for $|\mu_{eff}|^2\gg M_Z^2$ we have
\begin{equation}
|m_{\chi^0_1}|^2\le \displaystyle\frac{|\nu|^2\tilde{\sigma}^2}{\biggl(|\mu_{eff}|^2+\tilde{\sigma}^2+|\nu|^2\biggr)}\,.
\label{6}
\end{equation}
Thus in the considered limit the lightest neutralino mass is significantly smaller than $M_Z$ even for the
appreciable values of $\lambda$ at tree level.

When the mass of the lightest neutralino is small one can also obtain an approximate solution for $m_{\chi^0_1}$.
In general, the neutralino masses obey the characteristic equation $\mbox{det}\left(M_{\tilde{\chi}^0}-\kappa I\right)=0$,
where $\kappa$ is an eigenvalue of the matrix (\ref{1}). However, if $\kappa\to 0$ one can ignore all terms in this  
equation except the one which is linear with respect to $\kappa$ and
the $\kappa$--independent one which allows
to solve the characteristic equation. This method can be used to
calculate the mass of the lightest neutralino when $M_1, M_2$ and
$\mu_{eff}\gg M_Z$ because then the upper bound on $|m_{\chi^0_1}|$ 
goes to zero. 
We get in this limit (see \cite{neutralino,Hesselbach:2007ta})
\begin{equation}
|m_{\chi^0_1}|\simeq\displaystyle\frac{|\mu_{eff}|\nu^2\sin 2\beta}{\mu^2_{eff}+\nu^2}\,.
\label{7}
\end{equation} 
According to Eq.(\ref{7}) the mass of the lightest neutralino is inversely proportional to $\mu_{eff}$
and decreases when $\tan\beta$ grows. At small values of $\lambda$ the lightest neutralino mass is proportional 
to $\lambda^2$ because the correct breakdown of electroweak symmetry breaking requires $\mu_{eff}$ to remain constant 
when $\lambda$ goes to zero. At this point the approximate solution (\ref{7}) improves the theoretical restriction on 
the lightest neutralino mass derived above because for small values of $\lambda$ the upper bound (\ref{5})--(\ref{6}) 
implies that $|m_{\chi^0_1}| \propto \lambda$.
Note, however, that the lightest neutralino 
is predominantly singlino if $M_1, M_2$ and $\mu_{eff}\gg M_Z$
which makes its direct observation at future colliders quite challenging. 

\section*{Acknowledgments}
RN acknowledge support from the SHEFC grant HR03020 SUPA 36878.

\end{document}